# A Hybrid Agent Based and Differential Equation Model of Body Size Effects on Pathogen Replication and Immune System Response

Soumya Banerjee[1], and Melanie E. Moses[1]

[1]Dept. of Computer Science, University of New Mexico, Albuquerque, New Mexico
{soumya,melaniem}@cs.unm.edu

Many emerging pathogens infect multiple host species [1], and multi-host pathogens may have very different dynamics in different host species [2]. This research addresses how pathogen replication rates and Immune System (IS) response times are constrained by host body size. An Ordinary Differential Equation (ODE) model is used to show that pathogen replication rates decline with host body size but IS response rates remain invariant with body size. An Agent-Based Model (ABM) is used to investigate two models of IS architecture that could explain scale invariance of IS response rates. A *stage structured hybrid model* is proposed that strikes a balance between the detailed representation of an ABM and computational tractability of an ODE, by using them in the initial and latter stages of an infection, respectively.

The Immune System (IS) solves a search problem in both physical space and antigen space. The length of the search is determined by the time it takes for a cognate B-cell to encounter antigen. Our research suggests that this time is independent of the size of the organism [3]. This is counter-intuitive, since if we inject a sparrow and a horse with the same amount of antigen, the immune system of the horse has to search a larger physical space to find the pathogen, compared to the sparrow. This research attempts to explain how the time for the IS to search for antigen is independent of the size of the organism.

In addition to the immune system having to search larger spaces in larger organisms, larger body size can slow viral growth and immune system response times because the metabolic rate of cells is lower in larger species [4]. The metabolic rate of each cell is constrained by the rate at which nutrients and oxygen are supplied by the cardiovascular network. The rate at which this network supplies nutrients to each cell scales as the body mass ($M$) raised to an exponent of $-1/4$: $B_{cell} \propto M^{-1/4}$, such that individual cellular metabolic rates decrease as the body mass increases. The metabolic rate of a cell dictates the pace of many biological processes [4,5]. Metabolic rate is hypothesized to slow the speed of cell movement and proliferation in larger organisms. This could affect IS search times by reducing movement and proliferation of immune cells [6]. Rates of DNA and protein synthesis are also dependent on the cellular metabolic rate and could influence the rate at which pathogens replicate inside infected cells [2]. These two hypotheses, that IS search times and pathogen replication rates slow proportional to cellular metabolic rate and $M^{-1/4}$, lead to 4 possibilities, shown in Table 1 as originally proposed by Wiegel and Perelson [6].

**Table 1.** Four scaling hypotheses of pathogen replication and immune system response rate [6]

| H1: Pathogen replication rate $\propto M^0$ <br> IS search time $\propto M^0$ | H2: Pathogen replication rate $\propto M^{-1/4}$ <br> IS search time $\propto M^0$ |
|---|---|
| H3: Pathogen replication rate $\propto M^0$ <br> IS search time $\propto M^{-1/4}$ | H4: Pathogen replication rate $\propto M^{-1/4}$ <br> IS search time $\propto M^{-1/4}$ |

In the first test of the effects of body size on pathogen replication and immune system response rates, we combine a differential equation model, an Agent Based Model and empirical results from an experimental infection study [7]. The same West Nile Virus (WNV) strain was used to infect multiple avian species with body mass ranging from 0.03 kg (sparrows) to 3 kg (geese), and the viral load was monitored each day in blood serum over a span of 7 days post infection (d.p.i.)[7].

A standard Ordinary Differential Equation (ODE) model was used to simulate viral proliferation and immune response, and model results were compared to empirical levels of virus in blood [3]. In the model, $p$ = rate of virion production per infected cell, $\gamma$ = innate IS mediated virion clearance rate, $\omega$ = adaptive IS proliferation rate, $t_{pv}$ = time to attain peak viral load.

$$dT/dt = -\beta TV \qquad (1)$$

$$dI/dt = \beta TV - \delta I \qquad (2)$$

$$dV/dt = pI - c(t)V \qquad (3)$$

$$c(t) = \gamma,\ t < t_{pv} \qquad (4)$$

$$c(t) = \gamma e^{\omega(t - t_{pv})},\ t \geq t_{pv} \qquad (5)$$

Target cells $T$ are infected at a rate proportional to the product of their population and the population of virions $V$, with a constant of proportionality $\beta$. Infected cells $I$ die at a rate $\delta I$, and virions are cleared by the immune system at the rate $c(t)V$. The action of the immune system is decomposed into an innate response before peak viremia ($\gamma$), and an adaptive immune response after peak viremia characterized by a proliferation rate $\omega$. This paper focuses on $p$ as pathogen replication, and $\gamma$ and $\omega$ as IS response.

The parameters of the ODE model were fit to the viral load data for each of 25 species for days 1 – 7 d.p.i using non-linear least squares regression. According to the fits, pathogen replication rate ($p$, virions produced per infected cell per day) scaled as $p \propto M^{-0.29}$ (the predicted exponent of -0.25 is in the 95% CI, $r^2 = 0.31$, p-value = 0.0038). However, innate immune system mediated pathogen clearance rate ($\gamma$, day$^{-1}$) and adaptive immune system cell proliferation rate ($\omega$, day$^{-1}$) were independent of host mass $M$ (p-values of 0.4238 and 0.7242 respectively). These findings are consistent with hypothesis H2: pathogen replication rates decline in larger hosts, but immune response is independent of host mass.

Empirical data shows that time to peak viremia ($t_{pv}$) for WNV empirically occurs between 2 - 4 d.p.i., supporting the hypothesis that IS response rates are independent of $M$. If this peak were due to target cell limitation, then we would expect $t_{pv}$ to increase with host mass $M$, since the rate of pathogen replication decreases with $M$ and larger animals have more target cells. However, $t_{pv}$ could be determined by WNV specific antibodies, which have a critical role in WNV clearance [8]. If the peak is determined by a threshold presence of antibodies, it implies that the time for cognate B-cells to recognize antigen, proliferate and produce antibodies is independent of host mass: $t_{pv} = t_{detect} + t_{prolif} \propto M^0$.

These findings raise the question: what mechanisms make IS rates independent of host body mass and metabolism? A lymph node could have "privileged metabolism" which is independent of host mass [6]. Lymph nodes also form a *decentralized detection network* in which they are evenly distributed throughout the organism, and each lymph node serves as a central place in which IS cells, antigen presenting cells (e.g. dendritic cells) and antigen encounter each other in a small local region of tissue. This decentralized network could lead to efficient and expedient antigen detection that is independent of organism size.

The ODE model supports H2: for WNV, viral replication is constrained by host mass, but immune response appears independent of host mass. However, ODEs are unable to explain how the spatial arrangement of lymph nodes affects the time for the immune system to respond to infection.

## Two Competing Agent Based Models to Explore Mass Invariance of IS Response

In order to explore how the spatial arrangement of lymph nodes affects time to detect antigen, we used a spatially explicit Agent-Based Model (ABM). We used the CyCells [9] ABM to explicitly represent each cell and virion, and simulated viral replication in a 3D compartment representing the lymph node and draining tissue. The model is informed with data on Dendritic Cell (DC) and T-cell movement from recent *in-vivo* microscopy experiments [10]. We simulated DCs, B-cells, viruses and lymph nodes, and explicitly modeled DC migration from tissue to lymph node, and random walk of DC and B-cells in lymph node. We simulated DC morphology by letting them have a high surface area and large dendrite sweep area [10].

In our first model, we assumed that the lymphatic network forms a *decentralized detection network*, such that each lymph node and its draining tissue function as a unit of protection, which is iterated proportional to the size of the organism (similar to the concept of a *protecton* [11]), i.e. lymph nodes in all organisms are of the same size, lymph nodes are distributed evenly throughout the volume of each organism, and an organism 100 times bigger will have 100 times more lymph nodes, each of the same size. We also assumed that lymph nodes have *preferential metabolism* [6] i.e. inside a lymph node, IS cells have speed and proliferation rates that are invariant with host mass $M$. The ABM was then used to simulate a cubic compartment of length 2000μm, representing a lymph node and its draining tissue region. The model showed that the time taken for a cognate B-cell to detect antigen on a DC ($t_{detect}$) is independent of host mass and is slightly less than one day (20 simulations, mean = 16.43 hrs, SD = 13.83 hrs).

In the second model, we tested the alternative extreme that lymph nodes are arranged in a *centralized detection network* (bigger organisms have the same number of lymph nodes as smaller ones, however the size of an individual lymph node is larger) and they have preferential metabolism. We simulated 3 sizes of lymph nodes - a cubic lymph node of length 1000μm in a base animal, a cubic lymph node of length 2000μm in an animal 8 times bigger, and a cubic lymph node of length 5000μm in an animal 125 times bigger than the base animal. This model predicted that $t_{detect} \propto M^{0.93}$ (the exponent is theoretically predicted to be 1, and 1 is in the 95% CI, $r^2 = 0.99$, p-value = 0.01, 32 simulations) i.e. if a sparrow detects antigen in 12 hrs, then a goose would take 50 days, which is clearly unrealistic.

These results are consistent with our hypothesis that the decentralized nature of the lymphatic system, DC morphology and behavior, and privileged metabolism effectively reduce the antigen search time, so the classic search for a "needle in a haystack" problem can be solved in time independent of $M$. If $t_{pv} = t_{detect} + t_{prolif}$, and $t_{detect} \approx 1$ day independent of $M$, this implies that a fixed amount of time (1 to 3 days) is allocated to B-cell proliferation ($t_{prolif}$). Investigating how $t_{pv} = t_{detect} + t_{prolif}$ is independent of $M$, is ongoing.

## A Stage-Structured Hybrid Model

ODE models implicitly assume that populations are homogeneously mixed, for example, that at initialization, each injected virion has the opportunity to come in contact with every normal cell. This is unrealistic since inoculated virions localize at the site of infection. Such spatial effects assume more importance during the onset of infection, when the number of virions that are carried to the lymph nodes is small and depends on the spatial arrangement of lymph nodes and the spatial interactions of multiple cell types within the LN and draining tissue. An ABM can be used to model these complex interactions. However, due to the level of detail at which individual entities are represented, ABMs can be prohibitively computationally expensive. This study outlines an approach that aims to strike a balance between the detail of representation of an ABM and the computational tractability of an ODE model.

We call this a *stage-structured hybrid modeling* approach, which uses a detailed and spatially explicit, but computationally intensive Agent-Based Model (ABM) when spatial interactions matter, and a coarse-grained but computationally tractable Ordinary Differential Equation (ODE) model when the ODE assumptions of homogeneous mixture of population are likely to be satisfied and spatial effects can be ignored. We utilize this modeling approach to elucidate dependence of pathogen replication rates, and IS search times and rates, on host body size.

Our models support the hypothesis that pathogen replication rates decline with body mass, but IS detection is independent of mass. The ABM shows that the latter can be explained by the decentralized architecture of the lymphatic network.